\documentclass[11pt,a4paper]{article}
\usepackage[margin=26mm]{geometry}
\usepackage[numbers,sort&compress]{natbib}
\usepackage[T1]{fontenc}
\usepackage{lmodern}
\usepackage{amsmath,amssymb,amsthm,mathtools}
\usepackage{booktabs,array}
\usepackage{float}
\usepackage{tikz}
\usepackage[colorlinks=true,linkcolor=blue!45!black,citecolor=blue!45!black,urlcolor=blue!45!black]{hyperref}
\hypersetup{pdftitle={Locality Meets Interference: Structure and Scheduling in Radio Multicasting},pdfauthor={Qin Xin}}
\newtheorem{theorem}{Theorem}[section]
\newtheorem{lemma}[theorem]{Lemma}
\newtheorem{proposition}[theorem]{Proposition}
\newtheorem{corollary}[theorem]{Corollary}
\theoremstyle{definition}
\newtheorem{definition}[theorem]{Definition}
\newcommand{\dist}{\operatorname{dist}}
\newcommand{\rank}{\operatorname{rank}}
\newcommand{\ceil}[1]{\left\lceil #1\right\rceil}
\newcommand{\floor}[1]{\left\lfloor #1\right\rfloor}

\begin{document}

\title{Locality Meets Interference:\\Structure and Scheduling in Radio Multicasting}
\author{Qin Xin\\[4pt]\small Faculty of Science and Technology\\\small University of the Faroe Islands, T\'orshavn, Faroe Islands}
\date{}
\maketitle
\begin{abstract}
We study multipoint-to-multipoint multicasting in synchronous radio networks with unknown participant locations and unrestricted message aggregation. Let $n$ be the network size, $k$ the number of participants, and $d$ their hop diameter. A supplied distance-two coloring with $C$ colors yields completion within $(C-1)d+1$ slots and at most $nk$ transmissions. In unit disk graphs, coordinates, one-hop neighbor data, and a common degree bound $\Delta_0$ suffice for a $15(\Delta_0+1)$-color schedule. A known fixed separation bound permits optimal deterministic $\Theta(d)$ completion from coordinates alone. Without geometric data, strong selectivity gives $O(d\min\{n,(\Delta_0+1)^2\log n\})$ slots when $n$ is known. Independent random transmission gives $O((\Delta_0+1)(d+\log(k/\eta)))$ slots with failure probability at most $\eta$; for fixed $\Delta_0$, this simplifies to $O(d+\log(1/\eta))$. A star construction shows that the logarithmic term is necessary for this random rule when the degree grows. We extend color schedules to fixed deletions and subgraphs connected over bounded windows of cycles. Separate conditional reductions preserve additive costs under diameter guessing and control parameter-table and cluster preprocessing. Finally, a counterexample shows that pruning and reranking a breadth-first tree can destroy interference-free delivery. For one-hop designated delivery with receiver degree at most $b$, we prove that the optimal schedule length equals the chromatic number of an associated conflict graph and is at most $2b-1$, with matching examples.
\end{abstract}

\section{Introduction}
In multipoint-to-multipoint (M2M) multicasting, a set of $k$ participants exchanges its initial messages through an $n$-vertex radio network. Vertices outside the set may relay messages. The maximum distance $d$ between participants can be much smaller than the diameter $D$ of the whole network, so a bound in terms of $d$ captures the locality of the communication task.

The protocol must meet two requirements: move information toward participants whose locations are not known in advance, and arrange transmissions that their intended receivers can decode. Unlimited aggregation removes packet queues while leaving both requirements in place. We therefore state the structural inputs of each scheduling result explicitly, distinguishing a supplied coloring, coordinates, a degree bound, and full topology.

The destination set distinguishes several related tasks. Broadcasting delivers one source record to every vertex; gossiping delivers every vertex's record to every vertex. In $k$-gathering, the records of $k$ sources must reach a designated sink, whereas multi-broadcast delivers them to the whole network. M2M requires delivery only among the $k$ participants. These distinctions matter when $d\ll D$, and packet-size restrictions matter when many records share a route.

\subsection{Broadcasting and structural advice}
For known topology, Kowalski and Pelc~\cite{KP07} obtain deterministic broadcasting in $O(D+\log^2 n)$ rounds. Without topology knowledge, De Marco~\cite{DM10} gives a deterministic $O(n\log n\log\log n)$ bound even without a supplied network-size bound. Czumaj and Davies~\cite{CD18} subsequently develop universal and block synchronizers, obtaining improved parameter-sensitive deterministic broadcasting and wake-up guarantees when $n$, maximum in-degree, and source eccentricity are available. These results distinguish arbitrary identifiers from information installed by topology-aware preprocessing.

Randomization changes the dependence on network distance. Czumaj and Rytter~\cite{CR06} give broadcasting in $O(D\log(2n/D)+\log^2 n)$ rounds with high probability in unknown networks; for directed graphs, $D$ here denotes source eccentricity. We write $\log(2n/D)$ to keep the logarithmic factor bounded below by one. Our degree-based random rule in Section~\ref{sec:unknown} uses a fixed transmission probability and unrestricted aggregation. Its guarantee measures the time for that rule to exchange records among the participants.

Ga\'nczorz, Jurdzi\'nski and Pelc~\cite{GJP26} give deterministic broadcasting in $O(D+\log^2 n)$ rounds using constant-length labels, and gossiping in $O(D+\Delta\log n+\log^2 n)$ rounds using $O(\log\Delta)$-bit labels. These fastest bounds use existential labeling schemes; their polynomial-time constructive variants replace the additive $\log^2 n$ term by $\min\{D,\log n\}\log^2 n$. Their OPODIS 2025 paper was published in January 2026, following an October 2024 preprint. These results make the content and construction of advice essential comparison parameters. Our color protocol isolates what follows from one supplied distance-two coloring, with a bound in terms of $d$.

\subsection{Gossiping, gathering, and multicasting}
Known-topology gossiping has a substantial history~\cite{GPX04,GPX07,CMX06,CMX09}. In particular, G\k{a}sieniec, Potapov and Xin~\cite{GPX04} give an $n$-slot schedule and linear-diameter schedules for constant maximum degree. Section~\ref{sec:color} retains an elementary serial fallback and explains how the published $n$-slot result can instead be used.

For unknown strongly connected directed networks, the SWAT 2004 paper of G\k{a}sieniec, Radzik and Xin~\cite{GRX04} gives deterministic gossiping in $O(n^{4/3}\log^4 n)$ rounds with polynomially bounded identifiers. This improves the earlier $O(n^{5/3}\log^3 n)$ bound for polynomial identifiers and $O(n^{3/2}\log^2 n)$ bounds for linear identifiers. Its use of selective families establishes the $n^{4/3}$ benchmark against which later structural improvements are compared. Throughout this review, $\widetilde O$ suppresses polylogarithmic factors.

Wu and Chrobak~\cite{WC26} study sparse unknown directed networks. Their 2026 preprint reports $\widetilde O((mn)^{3/5})$ gossiping rounds for $m$ arcs and $\widetilde O(n\sqrt{\Delta})$ rounds for $\Delta$-regular graphs. The first bound improves the $\widetilde O(n^{4/3})$ benchmark when $m=O(n^c)$ for fixed $c<11/9$. These results concern all-to-all communication, with $k=n$, and complement our undirected participant-local setting. A numerical comparison alone would omit the different graph and information assumptions.

Xin~\cite{Xin26} studies the same unknown directed setting with polynomial identifiers, unrestricted messages, and no collision detection. Amplified reception certificates, adaptive pruning, and permutation selectors yield bounds sensitive to degree concentration and path covers. For common in- and out-degree $\Delta$, the reported bound is $\widetilde O(\min\{n\sqrt{\Delta},n^2/\Delta\})$; only $n$ and the identifier bound $N$ are protocol inputs. This structural approach complements the locality parameter $d$ studied here: its task disseminates all $n$ records globally, whereas M2M requires exchange only among the participants.

With message-sized packets, multi-message communication must also account for congestion. Ghaffari and Haeupler~\cite{GH13} construct collision-free layerings in
\[
 T_{\mathrm{lay}}=O\!\left(D\log(2n/D)+\log^{2+\varepsilon}n\right)
\]
rounds for every fixed $\varepsilon>0$. They obtain gathering of $k$ messages in $O(T_{\mathrm{lay}}+k)$ rounds, single-source $k$-message broadcasting in $O(T_{\mathrm{lay}}+k\log n)$ rounds, and gossiping in $O(n\log n)$ rounds, with high probability. Their packets carry one message-sized payload, with headers and coding where needed. This packet model differs from transporting an arbitrary collection of complete records in one slot, as permitted here.

Ga\'nczorz and Jurdzi\'nski~\cite{GJ26}, in SOFSEM 2026 with a December 2025 preprint, give $k$-gathering in at most $D+k$ rounds and multi-message $k$-broadcasting in $O(D+k+\log^2 n)$ rounds, using $O(\min\{\log k,\log\Delta\})$-bit advice. Their model permits nodes to wake only upon receiving a message. Comparison with M2M must account for oracle-assigned labels, the destination set, and the distinction between $D$ and $d$. An application inside Section~\ref{sec:reductions} would additionally require the stated clusters and valid local advice.

Krisko and Miller~\cite{KM26} characterize the advice needed for deterministic multi-broadcast. For $k,\Delta\ge2$, the tight worst-case label length is $\Theta(\min\{\log k,\log\Delta\})$ bits; they also characterize optimal gossiping labels for individual trees through the distinguishing number. Their 2026 journal article develops work first available in 2021. Its emphasis is feasibility and label length, rather than participant-local completion under a prescribed coloring.

Locality-sensitive M2M was studied by G\k{a}sieniec, Kranakis, Pelc and Xin~\cite{GKPX06} and by Xin, Manne, Zhang and Wang~\cite{XMW12}. Chlebus, Kowalski and Radzik~\cite{CKR09} distinguish many-to-many communication models by topology and participant information. The unit-disk $O(d+k)$ bound of~\cite{XMW12} uses known topology and participant parameters. Our coordinate-based bounds use different structural inputs. Their comparison therefore concerns both online slots and the information available before communication.

\subsection{Local broadcast, selectors, and energy}
Strongly selective families are established tools for isolating designated transmitters~\cite{CMS01}. Kuschner, Shashwat, Yadav and Chrobak~\cite{KSYC25} strengthen this perspective with permutation selectors: a probabilistic construction gives size $O(r^2\log N)$ while realizing every order on every $r$-element subset of an $N$-element identifier universe. Their ALGOWIN 2024 paper appears in a volume dated 2025; the preprint and online proceedings were available in 2024. The order property improves multi-hop gossiping schedules. Section~\ref{sec:unknown} needs only ordinary strong selectivity, so its existence bound is presented as a self-contained application of an established tool.

Garncarek, Jurdzinski, Kowalski, Kutten and Mosteiro~\cite{GJKKM25} study deterministic all-neighbor local broadcast in ISAAC 2025. Let $\gamma_2$ be the minimum, over dominating sets, of their maximum intersection size with a radius-two ball. They give $\widetilde O(\Delta\gamma_2^2)$-round algorithms trading advice length against adaptivity, and a lower bound
\[
 \Omega\!\left(\min\left\{\left(\frac{\min\{\Delta,\gamma_2\}}{\log n}\right)^2,n\right\}\right)
\]
for advice-free quasi-adaptive algorithms. This class permits local stopping of otherwise fixed transmission schedules. The distinction from Section~\ref{sec:ranks} is substantive: our designated-delivery schedule is constructed from known incidence data and has one required receiver per sender. The cited lower bound is for a different input and delivery requirement.

Energy is a separate objective from slot count or total transmissions. Banasik et al.~\cite{BDDGHP25}, in DISC 2025, give a randomized maximal independent set algorithm without collision detection using $O(\log^2 n\log\log n)$ awake rounds per vertex and $O(\log^3 n\log\Delta)$ rounds overall, with high probability. This result concerns the distributed construction of a coordination structure, illustrating why the cost of obtaining structural inputs should be accounted for separately from a protocol that receives them as advice.

Chang and Guan~\cite{CG26}, in SPAA 2026, give randomized aggregation without collision detection using $O(\log n)$-bit messages. With high probability, construction and execution take $O(n\operatorname{polylog}n)$ rounds and $O(\Delta^*\operatorname{polylog}n)$ energy, where $\Delta^*$ is the minimum possible maximum degree of a spanning tree. Their task computes a compact function of the inputs. Energy counts both sending and listening at the most active vertex; our $nk$ transmission bound concerns sending activity and does not bound listening energy.

\subsection{Contributions and organization}
Our first set of results isolates a common forwarding invariant. A distance-two coloring guarantees all-neighbor reception whenever a vertex transmits; forwarding an aggregate only after it changes then gives a locality-sensitive completion bound and at most $nk$ transmissions. We give two geometric ways to supply such a coloring. One-hop neighbor data permit a palette linear in a known degree bound, whereas a fixed separation bound permits a constant palette from coordinates alone. We then replace the coloring by either a strongly selective identifier schedule or independent random transmissions. The latter analysis tracks records along paths without assuming independent hop-progress indicators.

The remaining results identify which parts of this reasoning survive composition or structural changes. Color schedules tolerate deletions and connectivity over bounded windows of cycles, provided edges remain fixed within each cycle and unchanged records are retransmitted. For protocols that instead require parameter bounds, we give conditional reductions for diameter guessing, parameter-table discretization, and cluster composition. These reductions take their base protocols and structural inputs as premises; they do not establish the existence of cluster covers with arbitrary desired parameters. Finally, we examine interference directly. A pruning counterexample separates tree-rank bounds from valid radio schedules, while a conflict-graph characterization yields a tight $2b-1$ bound for one-hop designated delivery with receiver degree at most $b$.

Table~\ref{tab:bounds} summarizes the main completion guarantees and their inputs. Sections~\ref{sec:color}--\ref{sec:unknown} concern static networks; Section~\ref{sec:changes} treats changing subgraphs. Section~\ref{sec:reductions} gives the conditional reductions, and Section~\ref{sec:ranks} studies the interference constraints needed by local gathering schedules. Throughout, completion and termination are distinguished explicitly.

\begin{table}[t]
\centering\small
\renewcommand{\arraystretch}{1.08}
\caption{Completion bounds and their information requirements. Here $\Delta_0\ge\Delta$ is a common known degree bound, $0<\eta<1$, and $C_\epsilon$ is given by~\eqref{eq:palette}. The final row is a conditional reduction.}\label{tab:bounds}
\begin{tabular}{@{}>{\raggedright\arraybackslash}p{.29\linewidth}>{\raggedright\arraybackslash}p{.35\linewidth}>{\raggedright\arraybackslash}p{.30\linewidth}@{}}
\toprule
Available information or structure & Completion bound & Additional guarantee\\\midrule
Coloring of $G^2$ with $C$ colors & $(C-1)d+1$ & At most $nk$ transmissions\\
Known topology & $\Delta^2d+1$ & Polynomial preprocessing\\
Unit disk coordinates and one-hop neighbor data & $15(\Delta_0+1)d$ & Local computation of colors\\
Separated unit disk graph; coordinates and known $\epsilon$ & $C_\epsilon d$ & Deterministically optimal for fixed $\epsilon$\\
Known $n$, identifiers, and degree bound & $O(d\min\{n,(\Delta_0+1)^2\log n\})$ & Polynomial construction for fixed $\Delta_0$\\
Degree bound; independent random coins & $O((\Delta_0+1)(d+\log(k/\eta)))$ & Success probability at least $1-\eta$\\
Subgraphs of a fixed colored graph; union connected over $W$ cycles & $CW(n-1)$ & All records reach all vertices\\
Timed protocol with budget $A\delta+B$ & $5Ad+2B$ for unknown $d$ & Known $A,B$, cap $D_0\ge d$; resettable trials\\
\bottomrule
\end{tabular}
\end{table}

\section{Model and definitions}\label{sec:model}
Let $G=(V,E)$ be a finite simple connected undirected graph with $n=|V|\ge2$, maximum degree $\Delta$, and diameter $D$. Write $N(v)$ for the neighbors of $v$ and $N[v]=N(v)\cup\{v\}$. Vertices have distinct identifiers in $[n]=\{1,\ldots,n\}$. The participant set $S\subseteq V$ has $k\ge2$ members. Each $s\in S$ initially holds $(s,m_s)$, where $m_s$ is an arbitrary finite string known only to $s$. Nonparticipants initially hold no record. Set
\[
 d=\max_{s,t\in S}\dist_G(s,t),\qquad 1\le d\le D.
\]
Logarithms are base two unless written $\ln$. In asymptotic complexity bounds, logarithmic factors are truncated below at one; logarithms in exact identities and inequalities have their ordinary values. A degree bound $\Delta_0$, whenever used, is a common known integer satisfying $\Delta_0\ge\Delta$.

At the start of each synchronous slot, a vertex chooses its action and packet from its state and any fresh random coins. It either transmits or listens. A listener receives a packet at the slot's end exactly when one neighbor transmits; transmitters receive nothing, and silence is indistinguishable from collision. All vertices start together and share slot numbering. Packets may contain any finite collection of records and control information. An aggregate of all $k$ records costs one slot; with $b$-bit payloads, its explicit representation uses $O(k(b+\log n))$ bits. Local computation, preprocessing, and schedule installation are separate from online slot costs. Preprocessing uses only the structural information and parameters explicitly supplied in each theorem, with no additional participant-set or payload information.

Completion occurs when every participant holds all $k$ records; global termination also requires every vertex to halt. If $k$ is known, a participant recognizes completion by counting distinct source identifiers but continues the prescribed forwarding. Vertices remain active after local completion unless a stopping rule is given. Records are only copied and combined, singleton receptions are reliable, and no protocol uses collision detection.

The square $G^2$ connects distinct vertices at distance at most two in $G$. A proper coloring $c:V\to[C]$ of $G^2$ is a \emph{distance-two coloring}. The online color protocol uses only $c(v)$, $C$, the common clock, and the local initial record; known-topology preprocessing can compute and install these values.

A unit disk graph has an embedding $p:V\to\mathbb R^2$ in which distinct $u,v$ are adjacent exactly when $\|p(u)-p(v)\|\le1$. A separated unit disk graph additionally satisfies $\|p(u)-p(v)\|\ge\epsilon>0$ for all distinct vertices. Fixed separation means that $\epsilon$ is independent of $n$. Geometric algorithms use exact coordinates in a shared frame, with exact arithmetic and floor operations for cell indices. Coordinates, neighbor data, and colors are inputs when specified in a theorem. All graph diameters are measured in hops.

\section{Color cycles and transmission bounds}\label{sec:color}
We first assume that a distance-two coloring has been supplied. This isolates the forwarding argument from the way the schedule is constructed. Section~\ref{sec:geometry} supplies geometric colorings, while Section~\ref{sec:unknown} replaces coloring by identifier-based or random schedules.

A cycle consists of $C$ slots, numbered $1,\ldots,C$. Vertex $v$ can transmit only in slot $c(v)$. It stores its accumulated set $K_v$ of records and the set $L_v$ sent in its last transmission. Initially $K_v=\{(v,m_v)\}$ for $v\in S$ and $K_v=\varnothing$ otherwise; every $L_v$ is empty. In its color slot, $v$ transmits $K_v$ if $K_v\ne L_v$ and then sets $L_v:=K_v$. Otherwise it listens. It also listens in every other slot. A received packet is combined with $K_v$ by taking their union. The accumulated sets never lose records. The assignment to $L_v$ stores a snapshot: later updates of $K_v$ do not change $L_v$. Consequently $L_v\subseteq K_v$ throughout the execution.

\begin{lemma}\label{lem:safe}
Every transmission of the color-cycle protocol is received by all neighbors of its sender.
\end{lemma}
\begin{proof}
Adjacent vertices have different colors, so every neighbor of a transmitter listens in that slot. Two neighbors of the same listener are at distance at most two and therefore have different colors. Hence a listener has at most one transmitting neighbor in the slot. Suppressing a transmission cannot violate either property.
\end{proof}

\begin{theorem}\label{thm:color}
Given a distance-two coloring with $C$ colors, the color-cycle protocol completes M2M within $(C-1)d+1\le Cd$ slots. Over its entire execution it uses at most $nk$ transmissions. During its first $r$ cycles, for every integer $r\ge0$, it uses at most $n\min\{r,k\}$ transmissions. Neither $d$ nor $k$ is required by the forwarding rule. For every $C\ge3$ and $d\ge C-1$, the slot bound is attained by this protocol on a path using all $C$ colors and with two participating endpoints.
\end{theorem}
\begin{proof}
\emph{Progress by cycles.} Fix a source $s$. We prove by induction on $r$ that after $r$ cycles every vertex at distance at most $r$ from $s$ knows its record. The assertion is immediate for $r=0$. Suppose it holds for $r-1$, and let $uv$ be the last edge of a shortest path from $s$ to a vertex $v$ at distance $r$. At the start of cycle $r$, vertex $u$ knows the record. If $u$ has transmitted it before, then $v$ already received it by Lemma~\ref{lem:safe}. Otherwise the record belongs to $K_u\setminus L_u$, so $u$ transmits it in its color slot during cycle $r$. The same lemma ensures that $v$ receives it. Applying the induction to every ordered pair of participants proves completion within $d$ cycles.

\emph{Slot bound.} Fix a source-to-target path $v_0,\ldots,v_\ell$, where $1\le\ell\le d$. Let $t_i$ be the first slot in which $v_i$ transmits the source record, for $0\le i<\ell$. The source sends by its first color slot, so $t_0\le C$. Suppose $t_{i-1}$ is defined. By Lemma~\ref{lem:safe}, $v_i$ has the record at the end of that slot. If it sent the record earlier, then $t_i\le t_{i-1}$. Otherwise the record is absent from $L_{v_i}$, so $v_i$ sends it in its next color slot. Adjacent colors differ, and hence that slot occurs at most $C-1$ slots later. Inductively, all these times exist and satisfy
\[
 t_i\le t_{i-1}+C-1\quad(1\le i<\ell).
\]
The target receives the record by slot $t_{\ell-1}\le C+(\ell-1)(C-1)\le(C-1)d+1$. This proves the sharper bound, including the case $\ell=1$.

\emph{Tightness.} Take the path $v_0,\ldots,v_d$, make its endpoints the only participants, and assign
\[
 c(v_i)=1+((C-1-i)\bmod C).
\]
For $C\ge3$, vertices at distance one or two have different colors, so this is a valid distance-two coloring with palette $[C]$; all $C$ colors occur when $d\ge C-1$. The record of $v_0$ can reach $v_d$ only along the path. Its first transmission occurs in slot $C$, and at every subsequent relay the next color slot is exactly $C-1$ slots later. Thus $v_d$ first receives that record in slot $C+(d-1)(C-1)=(C-1)d+1$. The upper bound already proved makes this the exact completion time of the instance. This is a tightness statement for the supplied-color protocol, not a lower bound for arbitrary radio algorithms.

\emph{Transmission count.} Since $L_v\subseteq K_v$, every transmission by a vertex contains a record absent from its previous transmission. The corresponding sets $L_v$ form a strictly increasing sequence of subsets of the $k$ source records. Consequently each vertex transmits at most $k$ times. It also transmits at most once per cycle. Summing the two bounds over the vertices proves the transmission statements.
\end{proof}

The forwarding rule uses no participant count. Knowledge of $k$ supplies a stopping certificate: after forwarding all $k$ records, a vertex can safely halt.

\begin{corollary}\label{cor:stop}
If all vertices know $k$, each can halt permanently after first transmitting the complete set of records. Let $T_{\mathrm{halt},S}$ and $T_{\mathrm{halt},V}$ be the times by which all participants and all vertices, respectively, have halted. Then
\[
 T_{\mathrm{halt},S}\le(C-1)(d+1)+1,
 \qquad T_{\mathrm{halt},V}\le(C-1)(D+1)+1.
\]
In particular, the corresponding cycle bounds are $d+1$ and $D+1$. Neither diameter is required by the stopping rule. Both slot bounds are attained by the path family in Theorem~\ref{thm:color}.
\end{corollary}
\begin{proof}
When a vertex sends its final complete aggregate, each neighbor either has already halted with all records or is still active and listens in that slot. Every active neighbor receives the aggregate by the coloring property. Hence, whenever a vertex halts, all its neighbors know all records. A halted relay therefore cannot obstruct a record that its successor still lacks: it has already transmitted that record to the successor. The first-transmission argument in Theorem~\ref{thm:color} continues to apply, giving completion by slot $(C-1)d+1$ at every participant and by slot $(C-1)D+1$ at every vertex.

Because $k\ge2$, no vertex initially holds all records. Each vertex first completes its set at the end of a reception slot, which differs from its own color slot. Its last-sent snapshot is still incomplete, so its next color opportunity produces a transmission of the complete set. That opportunity occurs within $C-1$ slots, after which the vertex halts. Adding this delay gives both slot bounds. Since $(C-1)(h+1)+1\le C(h+1)$ for $h\ge1$, the cycle bounds follow as well.

For tightness, use the path and coloring from Theorem~\ref{thm:color}. The endpoint $v_d$ first receives the other participant's record in slot $(C-1)d+1$. Its next color slot is exactly $C-1$ slots later, when it first sends the complete set and halts. Here $D=d$, so both bounds are attained.
\end{proof}

We next translate the color bound into a known-topology guarantee. Its lower bound is the causal cost of moving a source record across $d$ edges.

\begin{corollary}\label{cor:degree}
Known-topology M2M admits polynomial preprocessing and completion within $\Delta^2d+1$ slots. For fixed maximum degree, the deterministic completion complexity is $\Theta(d)$ in the worst case over input records.
\end{corollary}
\begin{proof}
A vertex has at most $\Delta+\Delta(\Delta-1)=\Delta^2$ neighbors in $G^2$. Greedy coloring therefore uses at most $\Delta^2+1$ colors. The square graph and the coloring can be constructed in polynomial time, and Theorem~\ref{thm:color} applies.

For the lower bound, fix the graph, participants, and payload-independent preprocessing. Choose $s,t\in S$ at distance $d$. Induction on slots shows that after $j$ slots, the state of $t$ depends only on initial states within distance $j$: a state transition uses only the vertex's state and its neighbors' transmissions. Changing only $m_s$ therefore leaves the state of $t$ unchanged through slot $d-1$. Correct outputs must distinguish these payloads, so completion cannot be guaranteed before slot $d$.
\end{proof}

The next proposition identifies the palette cost of guaranteeing reception at every neighbor in each assigned slot.

\begin{proposition}\label{prop:palette}
A distance-two coloring of a graph of maximum degree $\Delta$ requires at least $\Delta+1$ colors. More generally, if every vertex is assigned one slot in a repeated schedule and every scheduled transmission must be received by every neighbor when all vertices transmit in their assigned slots, then the slot assignment is a proper coloring of $G^2$.
\end{proposition}
\begin{proof}
The closed neighborhood of a maximum-degree vertex is a clique of size $\Delta+1$ in $G^2$. For the second assertion, adjacent vertices cannot share a slot because a transmitter does not listen. Vertices with a common neighbor cannot share a slot because their packets collide there. These are exactly the pairs joined in $G^2$.
\end{proof}

When the degree is large, a network-size bound can be preferable to a color-based bound. A serial tree schedule supplies a fallback, and interleaving combines any two available protocols.

\begin{proposition}\label{prop:serial}
Known-topology M2M can be completed within $2(n-1)$ slots. If the inputs required by two correct protocols are available, they can be interleaved to obtain twice the smaller of their completion times.
\end{proposition}
\begin{proof}
Choose a rooted spanning tree. In a postorder of its nonroot vertices, let each vertex transmit its accumulated records once, using a separate slot for each transmission. Every child transmits before its parent, so the root obtains all records after $n-1$ slots. Then visit the nonleaf vertices in preorder, letting each transmit the complete aggregate in a separate slot. There are at most $n-1$ such vertices, and every child receives the complete aggregate before it needs to forward it. Only one vertex transmits in any slot, so other graph edges cause no collisions.

For interleaving, give the first protocol odd slots and the second even slots. Maintain separate states, original input copies, and a logical slot counter for each protocol. Each component controls only its allocated slots; halting one component leaves the other simulation active. Each execution therefore sees an exact simulation at half speed. Retain delivered records in an output buffer that does not alter either simulation. This proves the claimed minimum bound.
\end{proof}

The elementary serial bound above is useful without further scheduling machinery, but is not the best known global fallback. The $n$-slot gossiping theorem of~\cite{GPX04} also solves M2M: every vertex supplies a logical token, participants attach their genuine records, and other tokens are control placeholders discarded from the output. With that theorem available, interleaving yields completion within $2\min\{n,(C-1)d+1\}$ slots.

\section{Geometric schedules}\label{sec:geometry}
The generic square coloring uses $\Delta^2+1$ colors. Geometry permits a smaller palette by placing nearby vertices in cells and separating cells that reuse a color. We first allow multiple vertices in a cell and distinguish them by local identifier order; minimum separation will then remove that local-information requirement.

\begin{theorem}\label{thm:udg}
In a unit disk graph with a known common degree bound $\Delta_0$, a vertex's coordinates and its one-hop neighbors' identifiers and coordinates suffice to compute a distance-two coloring with $15(\Delta_0+1)$ colors. The resulting M2M completion bound is $15(\Delta_0+1)d$ slots.
\end{theorem}
\begin{proof}
Partition the plane into half-open square cells of side $s=1/\sqrt2$. Index the cells by $(i,j)\in\mathbb Z^2$, with cell $(i,j)$ equal to $[is,(i+1)s)\times[js,(j+1)s)$. Two points in a cell are at distance strictly less than one. Thus the vertices in each cell form a clique and number at most $\Delta_0+1$. Each vertex computes its position $a$ in the identifier order of its cell. Every other vertex in that cell is a one-hop neighbor, so the given local information suffices.

Assign the cell residue
\[
 \rho(i,j)=(i-4j)\bmod 15\in\{0,\ldots,14\},
\]
and give the vertex numerical color $c(v)=(\Delta_0+1)\rho(i,j)+a$. Equal numerical colors have equal residues and equal within-cell ranks. Vertices in the same cell therefore have distinct colors.

Consider two distinct cells with the same residue, and write $(u,v)$ for the difference of their indices. If $|u|\ge4$ or $|v|\ge4$, the corresponding coordinate gap between any two points in the cells is greater than $3s>2$. Otherwise $|u|,|v|\le3$ and $u-4v\equiv0\pmod{15}$. Since $|u-4v|\le15$, we have $u-4v\in\{-15,0,15\}$. Applying the coordinate bounds to these three equations gives exactly $(u,v)=(0,0),(3,-3),(-3,3)$. The first is excluded because the cells differ. In either remaining case, both coordinate gaps are strictly greater than $2s$, using the half-open cell convention. The Euclidean distance is then greater than $\sqrt{(2s)^2+(2s)^2}=2$.

Thus equal-colored distinct vertices are more than two Euclidean units apart and cannot be joined by two unit-length edges. The coloring is proper for $G^2$ and uses at most $15(\Delta_0+1)$ colors. The common degree bound fixes the palette and slot order. Theorem~\ref{thm:color} gives the claimed completion time and, more precisely, $[15(\Delta_0+1)-1]d+1$ slots.
\end{proof}

For $\Delta_0=\Delta$, Proposition~\ref{prop:palette} shows that this palette size is within a factor of fifteen of the smallest possible distance-two palette. This comparison concerns the number of colors, rather than a lower bound on arbitrary M2M protocols. With full topology one may choose the smaller of the geometric and greedy palettes.

\begin{theorem}\label{thm:separated}
In a separated unit disk graph, coordinates and a common known separation bound $\epsilon$ suffice to assign a distance-two coloring with
\begin{equation}\label{eq:palette}
 C_\epsilon=\left(\ceil{2\sqrt2/\epsilon}+1\right)^2
\end{equation}
colors. No neighbor information is required. Completion takes at most $C_\epsilon d$ slots, and for fixed $\epsilon$ the worst-case deterministic completion complexity is $\Theta(d)$.
\end{theorem}
\begin{proof}
Use half-open square cells of side $s=\epsilon/\sqrt2$. A cell contains at most one vertex, because the distance between two points in it is strictly less than $\epsilon$. Let $M=\ceil{2/s}+1$, and assign color $1+M(i\bmod M)+(j\bmod M)$ to the vertex in cell $(i,j)$, with residues in $\{0,\ldots,M-1\}$. For distinct cells of one color, at least one cell index differs by a nonzero multiple of $M$. In that coordinate, every pair of points has a gap strictly greater than $(M-1)s\ge2$. The strict inequality follows from the half-open upper cell boundary and also covers the case where $2/s$ is an integer. Equal-colored vertices cannot be joined by a path of at most two edges. The number of colors is $M^2=C_\epsilon$. The upper bound follows from Theorem~\ref{thm:color}. The causal lower-bound argument in Corollary~\ref{cor:degree} still applies when coordinates and a separation bound are supplied, since these inputs contain no source payload. It gives the matching $\Omega(d)$ lower bound.
\end{proof}

For example, $\epsilon=1/2$ gives $M=7$ and $C_\epsilon=49$. Separation also supplies a direct degree bound, independently of the grid construction.

\begin{lemma}\label{lem:packing}
For every vertex $v$ in a separated unit disk graph and every integer $h\ge0$, the number of vertices within $h$ graph hops of $v$ is at most $(1+2h/\epsilon)^2$. In particular,
\[
 \Delta+1\le(1+2/\epsilon)^2.
\]
\end{lemma}
\begin{proof}
Vertices within $h$ hops of $v$ lie within Euclidean distance $h$ of $p(v)$. Place an open disk of radius $\epsilon/2$ at each such vertex. The disks have disjoint interiors by separation and lie in the disk of radius $h+\epsilon/2$ centered at $p(v)$. Comparing areas gives the result. For $h=1$, choose a maximum-degree vertex.
\end{proof}

Because a connected unit disk graph with at least two vertices has $\epsilon\le1$, equation~\eqref{eq:palette} also gives $C_\epsilon=O(\epsilon^{-2})$. The online color rule remains the same as in Section~\ref{sec:color}, so all geometric bounds retain the $nk$ transmission guarantee.

\section{Protocols for unknown topology}\label{sec:unknown}
When a distance-two coloring and geometric inputs are unavailable, an algorithm can instead arrange isolated receptions directly. We give two alternatives: a deterministic schedule that works for every neighborhood of bounded size, and a random rule whose progress is analyzed along paths. Both use unrestricted aggregation.

\subsection{A deterministic selective schedule}
\begin{definition}
A family $\mathcal F\subseteq2^{[n]}$ is \emph{strongly $r$-selective} if for every $Z\subseteq[n]$ with $1\le|Z|\le r$ and every $z\in Z$, some $F\in\mathcal F$ satisfies $F\cap Z=\{z\}$.
\end{definition}

\begin{lemma}\label{lem:selective}
For $2\le r\le n$ there exists a strongly $r$-selective family of size $O(\min\{n,r^2\log n\})$. For every fixed $r$, such a family can be constructed deterministically in polynomial time in $n$.
\end{lemma}
\begin{proof}
Generate $T$ rows, including each identifier independently with probability $1/r$ in each row. A row isolates a specified pair $(Z,z)$ with probability
\[
 \frac1r(1-1/r)^{|Z|-1}\ge\frac1r(1-1/r)^{r-1}\ge\frac1{er}.
\]
The last inequality follows from $(1+1/(r-1))^{r-1}\le e$. The number of pairs is
\[
 \sum_{s=1}^r s\binom ns
 =n\sum_{s=1}^r\binom{n-1}{s-1}\le r n^r.
\]
Consequently the expected number of pairs isolated by no row is at most $r n^r\exp(-T/(er))$. Taking $T=4r(r+1)\ceil{\log n}$ makes this expectation less than one, because
\[
 \ln r+r\ln n\le(r+1)\ln n
 <\frac4e(r+1)\ceil{\log n}=\frac{T}{er}.
\]
Thus $T=O(r^2\log n)$ rows suffice.

For the deterministic construction, fix lexicographic orders on all pairs $(Z,z)$ and on the $nT$ membership bits, then expose the bits in that order. For each value of the next bit, compute the conditional expected number of unisolated pairs. For a given pair, its failure probability is the product over rows of one minus that row's conditional isolation probability. Each row probability is computed from the fixed bits and remaining independent membership probabilities. The expectation before fixing a bit is a convex combination of the expectations for its two values, so some value does not increase it. Choose such a value, using a fixed rule to break ties.

When all bits have been fixed, the number of unisolated pairs is an integer below one and hence is zero. Discarding duplicate rows gives the required family. For fixed $r$, there are polynomially many pairs and bits. All computed probabilities are rational numbers with polynomial bit length: their denominators divide a power of $r$ with exponent at most $rT$. Thus the construction, including its arithmetic, takes polynomial time for fixed $r$.

Finally, the $n$ singleton subsets form a strongly $r$-selective family for every $r\le n$. If the prescribed row count $T$ is at least $n$, use this explicit family instead; otherwise use the construction above. This proves the stated minimum without changing the fixed-$r$ polynomial-time guarantee.
\end{proof}

When $n$ is known, any larger supplied degree bound may first be replaced by $n-1$. We can therefore use the following parameter range without losing generality.

\begin{theorem}\label{thm:detunknown}
Suppose every vertex knows $n$, its identifier, and a common integer degree bound $\Delta_0$ with $\Delta\le\Delta_0\le n-1$. A topology-independent schedule completes M2M in $O(d\min\{n,(\Delta_0+1)^2\log n\})$ slots. Its offline construction is polynomial for fixed $\Delta_0$; in that case the completion bound is $O(d\log n)$.
\end{theorem}
\begin{proof}
Use Lemma~\ref{lem:selective} with $r=\Delta_0+1$. All vertices use the same deterministically chosen family and ordering of its rows. One cycle consists of these rows. Each vertex freezes a copy of its accumulated records at the start of the cycle. In row $F$, the vertices with identifiers in $F$ transmit their frozen copies, including empty packets, and the others listen. Received records are accumulated for later cycles.

For an oriented edge $uv$, take $Z=N[v]$ and $z=u$. Since $|N[v]|\le\Delta_0+1$, a row isolates $u$ in $Z$. In that row $u$ transmits, $v$ listens, and all other neighbors of $v$ listen. Every record known by $u$ at the cycle's start therefore reaches $v$ during the cycle. By induction on distance, every source record reaches every participant after at most $d$ cycles. Multiplying by the family size proves the bound.
\end{proof}

\subsection{Independent random transmission}
The selective schedule bounds the time for each hop deterministically. For random transmission, we instead track one copy of a record along a fixed path and bound its chance of advancing in each slot. Its position depends on earlier slots, so the analysis needs a tail bound that allows this dependence.

\begin{lemma}\label{lem:progress}
Let $(\mathcal H_t)_{t=0}^T$ be a filtration and let $X_t\in\{0,1\}$ be $\mathcal H_t$-measurable. Suppose $0<q\le1$ and $\Pr(X_t=1\mid\mathcal H_{t-1})\ge q$ almost surely for every $1\le t\le T$. Put $\mu=Tq$. Then
\[
 \Pr\!\left[\sum_{t=1}^T X_t<\mu/2\right]\le e^{-\mu/8}.
\]
\end{lemma}
\begin{proof}
For every $t$, almost surely,
\[
 \mathbb E[e^{-X_t}\mid\mathcal H_{t-1}]
 \le 1-q+q/e\le\exp[-q(1-e^{-1})].
\]
Since $\sum_{i<t}X_i$ is $\mathcal H_{t-1}$-measurable, iterating conditional expectation gives
$\mathbb E[\exp(-\sum_tX_t)]\le\exp[-\mu(1-e^{-1})]$.
Markov's inequality now bounds the stated probability by
\[
 e^{\mu/2}\mathbb E[e^{-\sum_tX_t}]
 \le\exp[-\mu(1/2-e^{-1})]\le e^{-\mu/8},
\]
where $e^{-1}<3/8$. This argument does not assume mutual independence of the $X_t$.
\end{proof}

\begin{theorem}\label{thm:random}
Let a common integer $\Delta_0\ge\Delta$ be known. In each slot, every vertex independently transmits its current set of records with probability $p=1/(\Delta_0+1)$ and otherwise listens. Fresh coins are used in every slot, and empty packets may be transmitted. For every $0<\eta<1$, M2M completes with probability at least $1-\eta$ within
\[
 O\!\left((\Delta_0+1)\left[d+\log(k/\eta)\right]\right)
\]
slots. The rule runs continuously using only the degree bound and fresh random coins. The parameters $k$, $d$, and $\eta$ enter only the completion guarantee; no stopping time is asserted.
\end{theorem}
\begin{proof}
Fix distinct participants $s,t$ and a shortest path $v_0=s,v_1,\ldots,v_\ell=t$, where $\ell\le d$. Track a marked copy of the record along this path. The mark starts at $v_0$ and advances by at most one edge per slot. If it is currently at $v_i$, an advance occurs when $v_i$ transmits and all vertices in $N[v_{i+1}]\setminus\{v_i\}$ listen. The marked record then reaches $v_{i+1}$. Other successful receptions can only make the actual dissemination earlier than this tracked process.

Before it reaches $t$, the marked location is determined by the previous history. Conditional on that history, the probability of an advance is
\[
 p(1-p)^{\deg(v_{i+1})}\ge p(1-p)^{\Delta_0}
 =\frac{(1+1/\Delta_0)^{-\Delta_0}}{\Delta_0+1}
 \ge q:=\frac1{e(\Delta_0+1)}.
\]
Here $(1+1/\Delta_0)^{\Delta_0}\le e$, and fresh slot coins are independent of the preceding history. Use the filtration $\mathcal H_j$ generated by the execution through slot $j$. Let $X_j$ indicate an advance in slot $j$, and set $X_j=1$ in every slot after the mark arrives. The variables $X_j$ are $\mathcal H_j$-measurable, and the dummy successes ensure that $\Pr(X_j=1\mid\mathcal H_{j-1})\ge q$ for every slot. If the mark has not arrived by time $T$, then $\sum_{j=1}^T X_j<\ell$.

Set $a=\ln(k^2/\eta)$ and $T=\ceil{8(d+a)/q}$. Then $\mu=Tq\ge8(d+a)$ and $\ell\le d\le\mu/2$. Failure of actual delivery implies that the mark has not arrived, and hence that $\sum_jX_j<\ell\le\mu/2$. Lemma~\ref{lem:progress} therefore gives
\[
 \Pr[\text{the record from $s$ has not reached $t$ by $T$}]
 \le e^{-\mu/8}\le\eta/k^2.
\]
A union bound over the $k(k-1)$ ordered pairs proves the success guarantee. Since $q^{-1}=e(\Delta_0+1)$ and $\ln(k^2/\eta)\le2\ln(k/\eta)$, the stated bound follows. All records share aggregate packets, so they incur no queueing delay. The paths are used only in the analysis.
\end{proof}

The high-probability bound has two immediate consequences. The first uses the fact that bounded degree also limits the number of participants in a ball of radius $d$; the second concerns completion time even though the protocol does not detect completion.

\begin{corollary}\label{cor:randomfixed}
For fixed $\Delta_0$ and every $0<\eta<1$, the random-transmission protocol completes in $O(d+\log(1/\eta))$ slots with probability at least $1-\eta$. In particular, for fixed $c>0$ it completes in $O(d+\log n)$ slots with probability at least $1-n^{-c}$.
\end{corollary}
\begin{proof}
Fix a participant $s$. Every participant belongs to its radius-$d$ ball. Counting walks of lengths at most $d$ gives
\[
 k\le\sum_{i=0}^{d}\Delta_0^i\le(\Delta_0+1)^d.
\]
The second inequality follows from the binomial expansion, since $d\ge1$. Hence $\log k\le d\log(\Delta_0+1)$. Substitute into Theorem~\ref{thm:random}; for fixed $\Delta_0$ this proves the first assertion. Take $\eta=n^{-c}$ for the second.
\end{proof}

\begin{corollary}\label{cor:randommean}
The random-transmission protocol completes almost surely. Its completion time $T_{\mathrm{M2M}}$ satisfies
\[
 \mathbb E[T_{\mathrm{M2M}}]=O((\Delta_0+1)(d+\log k)).
\]
For fixed $\Delta_0$, the expectation is $\Theta(d)$. These are completion-time bounds, not bounds on a locally detected stopping time.
\end{corollary}
\begin{proof}
The proof of Theorem~\ref{thm:random} gives, for every integer $T$ with $Tq\ge2d$,
\[
 \Pr(T_{\mathrm{M2M}}>T)\le\min\{1,k(k-1)e^{-qT/8}\},
 \qquad q=\frac1{e(\Delta_0+1)}.
\]
The right side tends to zero, proving almost-sure completion. Put $T_0=\ceil{(8/q)(d+2\ln k)}$. For every integer $u\ge0$, the same bound gives $\Pr(T_{\mathrm{M2M}}>T_0+u)\le e^{-qu/8}$. Summing the tail yields
\[
 \mathbb E[T_{\mathrm{M2M}}]\le T_0+\sum_{u\ge0}e^{-qu/8}
 =T_0+\frac1{1-e^{-q/8}}
 \le T_0+16/q
 =O(q^{-1}(d+\log k)),
\]
where $1-e^{-q/8}\ge q/16$ for $0<q\le1$. For fixed $\Delta_0$, the ball-counting inequality in Corollary~\ref{cor:randomfixed} gives $\log k=O(d)$, proving an $O(d)$ expectation. A record can cross at most one edge per slot, so a pair of participants at distance $d$ cannot complete its exchange earlier than slot $d$ under this forwarding rule. Hence the expectation is also at least $d$. No stopping rule is used.
\end{proof}

The fixed-degree conclusion uses the restriction that only exponentially many participants fit within distance $d$. As the degree grows, even a participant set of diameter two can incur a logarithmic collection cost. The next star example gives a matching lower bound for the independent random rule of Theorem~\ref{thm:random}.

\begin{proposition}\label{prop:randomstar}
On a star with $L\ge2$ participating leaves and a nonparticipant center, use the random-transmission rule with $\Delta_0=L$. Then
\[
 \mathbb E[T_{\mathrm{M2M}}]=\Theta(L\log L).
\]
For $0<\eta\le1/2$, achieving completion probability at least $1-\eta$ requires $\Omega(L\log(L/\eta))$ slots; the upper bound of Theorem~\ref{thm:random} has the same order on this instance.
\end{proposition}
\begin{proof}
Here $k=L$ and $d=2$. In a slot, a specified leaf succeeds at the center with probability
\[
 r=p(1-p)^L,\qquad p=\frac1{L+1},\qquad r=\Theta(1/L).
\]
The exponent counts the center and the other $L-1$ leaves. Every route between distinct leaves passes through the center. Thus a record not yet received by the center is present only at its source leaf, and its first reception must be directly from that leaf. The successful sender identity is independent between slots, and successes of different leaves are disjoint within a slot. If $i$ leaf records have not reached the center, the probability of receiving a new one in the next slot is $ir$. The expected time to receive all of them is therefore $H_L/r=\Theta(L\log L)$, where $H_L=\sum_{i=1}^L1/i$. This lower-bounds M2M completion; Corollary~\ref{cor:randommean} supplies the upper bound.

For the confidence bound, let $M$ count leaf records never received by the center in the first $T$ slots. Then $m:=\mathbb E[M]=L(1-r)^T$. For two different leaves, the probability that neither succeeds is $(1-2r)^T\le(1-r)^{2T}$, so $\operatorname{Var}(M)\le m$. Cauchy--Schwarz gives
\[
 \Pr(M>0)\ge\frac{(\mathbb E M)^2}{\mathbb E[M^2]}
 \ge\frac{m}{1+m}.
\]
If completion has probability at least $1-\eta$, then $\Pr(M>0)\le\eta$ and $m\le\eta/(1-\eta)$. Consequently
\[
 T\ge\frac{\ln(L(1-\eta)/\eta)}{-\ln(1-r)}
 =\Omega(L\log(L/\eta)),
\]
Specifically, $r\le1/3$ gives $-\ln(1-r)\le3r/2$, while $L\ge2$ and $\eta\le1/2$ give $\ln(L(1-\eta)/\eta)\ge\tfrac12\ln(L/\eta)$. Together with $r=\Theta(1/L)$ these inequalities justify the displayed asymptotic lower bound uniformly in $L$ and $\eta$. Theorem~\ref{thm:random} gives the matching upper bound because $d=2$.
\end{proof}

\section{Deletions and changing subgraphs}\label{sec:changes}
The static proofs separate collision avoidance from progress. Deleting edges preserves the coloring and therefore preserves collision avoidance. Progress must instead be measured in the surviving graph. When edges change repeatedly, a previously sent record may need to cross a newly available edge, so forwarding only after a state change is no longer sufficient.

\begin{proposition}\label{prop:deletions}
Let $G_0$ have an installed distance-two coloring with $C$ colors, and let the operating graph $H$ be a fixed connected subgraph obtained by vertex and edge deletions before the protocol starts. Initialize each surviving participant with its original record. If at least two participants survive, the color-cycle protocol completes their exchange within $(C-1)d_H+1$ slots, where $d_H$ is their diameter in $H$. It uses at most $|V(H)|\,|S\cap V(H)|$ transmissions.
\end{proposition}
\begin{proof}
Every path of at most two edges in $H$ is also a path in $G_0$. The restricted coloring is therefore proper for $H^2$. Apply Theorem~\ref{thm:color} in $H$ with the surviving participants and their records. Its proof uses only identifier distinctness, so the original identifiers can be retained. The forwarding rule needs no participant count; if the stopping rule of Corollary~\ref{cor:stop} is used, the count must be that of the surviving participants.
\end{proof}

For changes between cycles, use the fixed coloring but transmit in every assigned slot. This lets a newly available edge carry records sent earlier.

\begin{theorem}\label{thm:dynamic}
Let $G_0=(V,E_0)$ have a distance-two coloring with $C$ colors. In cycle $i$, let the operating graph be a spanning subgraph $H_i\subseteq G_0$, fixed throughout that cycle. Suppose that for some integer $W\ge1$ the union of every $W$ consecutive cycle graphs is connected. If every vertex broadcasts its current accumulated records in its color slot in every cycle, then every initial record reaches every vertex within $CW(n-1)$ slots. The graphs $H_i$ may depend on the preceding execution history, subject to the stated conditions. Neither $W$ nor the connectivity window is used by the forwarding rule.
\end{theorem}
\begin{proof}
The coloring remains proper for $H_i^2$, so every transmission reaches all current neighbors. Fix one source record and partition time into blocks of $W$ cycles. Let $I$ be its informed set at the beginning of one block. If $I\ne V$, the connected union of the block's cycle graphs contains an edge $uv$ with $u\in I$ and $v\notin I$. In a cycle containing that edge, $u$ still knows the record and transmits it in its color slot. Vertex $v$ either already knows it or receives it then. Thus the informed set grows by at least one vertex during each unfinished block. It starts nonempty and is monotone, so at most $n-1$ blocks suffice. The same argument holds simultaneously for all records.
\end{proof}

This protocol makes $n$ transmissions per cycle, so its stated deadline uses $nW(n-1)$ transmissions. The rule continues after completion unless an additional stopping rule is supplied. For $W=1$, every cycle graph is connected and the deadline is $C(n-1)$ slots. Stability within a cycle is a separate assumption, not a consequence of connectivity. For example, take a triangle host with colors $1,2,3$ and track a record initially at vertex $1$. In vertex $1$'s slots use the path $1$--$2$--$3$; in vertex $2$'s slots use $2$--$1$--$3$; in vertex $3$'s slots use any connected spanning graph. Vertex $2$ learns the record, but vertex $3$ never receives it. Every slot graph is connected, yet the edges change before the needed color slot. This example explains why the theorem fixes each operating graph for a whole cycle.

\section{Diameter guessing and preprocessing reductions}\label{sec:reductions}
The preceding forwarding protocols run without knowing $d$. The following reductions address protocols prepared for supplied parameter bounds or clusters. Each reduction takes its stated base protocols, parameter budgets, and structural data as inputs. These hypotheses are not consequences of the static-network theorems and must be verified in each application. As in Proposition~\ref{prop:serial}, combined executions use separate working states and clocks, while delivered records are retained in an output buffer.

\subsection{Preserving an additive term when the diameter is unknown}
Starting diameter trials at one can repeatedly charge an additive cost that dominates the early trials. Starting near the scale at which the diameter term balances that additive cost avoids this accumulation.

\begin{theorem}\label{thm:doubling}
Fix all parameters except participant diameter, and let a common integer cap $D_0\ge d$ be known. Suppose that for every integer $1\le\delta\le D_0$ a deterministic protocol $\mathcal P(\delta)$ completes whenever $d\le\delta$. Assume that a known sufficient budget is at most $A\delta+B$, where $A\ge1$ and $B\ge0$ are common known coefficients independent of $\delta$. Trials can be interrupted at their budgets, even when $\delta<d$, and reset to their initial working states while retaining the original source records. No trial fabricates records. Then M2M can be completed within $5Ad+2B=O(Ad+B)$ slots without knowledge of $d$.
\end{theorem}
\begin{proof}
Set
\[
 \delta_0=\min\{D_0,\max\{1,\ceil{B/A}\}\},
 \qquad \delta_i=\min\{D_0,2\delta_{i-1}\}.
\]
Run the trials consecutively, assigning trial $i$ exactly $\ceil{A\delta_i+B}$ slots. At its common starting boundary, reset every vertex's working state and set the trial's local slot counter to zero. A vertex whose component execution halts early sends nothing until the next boundary. Keep received output records in a separate monotone buffer that does not alter a fresh trial's state transitions. Local completion does not cause a vertex to abandon the prescribed trial execution.

Let $j$ be the first index with $\delta_j\ge d$; it exists because the guesses reach $D_0$. This index is used only in the analysis, and earlier trials may already have delivered some records. Since $\delta_0\le1+B/A$, the case $j=0$ gives
\[
 \ceil{A\delta_0+B}\le A+2B+1\le2(Ad+B)\le5Ad+2B.
\]
Suppose $j\ge1$. The initial guess is not clipped at $D_0$, so $B\le A\delta_0$. Before the final trial the guesses double, and clipping can only decrease the last guess. Therefore
\[
 \sum_{i=0}^{j}\delta_i\le(2^{j+1}-1)\delta_0,
 \qquad d\ge2^{j-1}\delta_0+1,
\]
where the second inequality uses integer diameters and $\delta_{j-1}<d$. The total rounded budget is at most
\[
 U:=A(2^{j+1}-1)\delta_0+(j+1)B+(j+1).
\]
Using the preceding inequalities, $A\ge1$, and $\delta_0\ge1$, we obtain
\begin{align*}
 5Ad+2B-U
 &\ge A\delta_0(2^{j-1}+1)-(j-1)B+5A-(j+1)\\
 &\ge A\delta_0(2^{j-1}-j+2)+5A-(j+1)\\
 &\ge 2^{j-1}-2j+6\ge0.
\end{align*}
For the last step, the values at $j=1,2,3$ are $5,4,4$, and thereafter the expression increases; also $2^{j-1}-j+2>0$. Hence the total budget is at most $5Ad+2B$. Trial $j$ succeeds by assumption. Resetting prevents earlier trials from changing its execution, and the output buffer preserves all genuine records delivered by any trial.
\end{proof}

The theorem bounds completion. Continuing through the trial at $D_0$ also gives a finite protocol with a common final stopping time. The bound on that final stopping time may be larger. One may take $D_0=n-1$ when $n$ is known and the base family is defined for every guess in $[1,n-1]$.

\subsection{Geometric discretization of parameter tables}
Diameter guessing controls online completion time. We next consider the separate offline cost of preparing a family of protocols. If each protocol accepts parameter upper bounds, it suffices to prepare a geometric grid of those bounds rather than every integer pair.

\begin{theorem}\label{thm:table}
Suppose $n$ and $D$ are supplied to preprocessing, and a protocol family accepts upper bounds $(\delta,\kappa)$ on participant diameter and count. Each member must be correct for every participant set with diameter at most $\delta$ and size at most $\kappa$, and must have preprocessing cost at most $P(n)$. Let $F(n,\delta,\kappa)$ be an upper bound on online time. Suppose there is a constant $c$ such that for all valid parameter pairs in $[1,D]\times[2,n]$,
\[
 F(n,\delta',\kappa')\le cF(n,\delta,\kappa)
 \quad\text{whenever}\quad
 \delta\le\delta'\le2\delta,\quad \kappa\le\kappa'\le2\kappa.
\]
Preparing only geometric grids of $\delta\in[1,D]$ and $\kappa\in[2,n]$ costs $O(P(n)\log(D+1)\log n)$ and preserves the online bound up to the factor $c$, when the requested parameter bounds are rounded to those grids.
\end{theorem}
\begin{proof}
For the diameter grid use the powers of two not exceeding $D$, together with $D$. For the count grid use the powers of two in $[2,n]$, together with $n$. Each requested integer parameter rounds upward to a grid point less than twice its value, or to itself. The grids contain $O(\log(D+1))$ and $O(\log n)$ values, respectively. Preprocess every pair. Correctness follows because the rounded values are upper bounds, and the displayed stability condition bounds the online cost. Multiplying the number of pairs by $P(n)$ proves the preprocessing bound.
\end{proof}

A routine valid for a count upper bound must also finish when fewer participants are present; waiting for exactly $\kappa$ distinct records does not satisfy this premise. For example, $P(n)=O(n^3)$ yields $O(n^3\log(D+1)\log n)$ total preprocessing under the theorem's hypotheses. The requested bounds are inputs to the selection of a table entry. In particular, this preprocessing reduction does not by itself give an online protocol for discovering an unknown participant count.

\subsection{Composing cluster schedules}
The geometric grid controls how many parameter choices are prepared. To use such protocols inside clusters, we must also control interference between concurrent executions. The following composition theorem states that extra requirement explicitly. A cluster may contain zero or one participant; its exchange is then complete initially, but its execution must still obey the prescribed slot schedule. The intrinsic diameter of a connected cluster $U\subseteq V$ is the diameter of its induced subgraph $G[U]$, so paths used by its local protocol remain inside the cluster.

\begin{theorem}\label{thm:cover}
Fix a participant-diameter bound $\delta$. Suppose a supplied family of connected clusters, partitioned into $\beta$ classes, has the following properties: every cluster has intrinsic diameter at most $\alpha\delta$; clusters in a common class are vertex-disjoint with no edge between them; and every participant set of size at most $k$ and diameter at most $\delta$ is contained in some cluster. Cluster memberships, class labels, and the protocol's required structural inputs are supplied before the online execution. Suppose a within-cluster protocol is defined for every local participant set of size at most $k$, is interruptible at class boundaries, and completes every such exchange within a supplied common integer budget $\tau=O(\alpha\delta+Q(n,k))$ slots. Then M2M for every such participant set can be completed in
\[
 O\!\left(\beta[\alpha\delta+Q(n,k)]\right)
\]
slots. If all vertices know the exact global count $k$, it suffices that the local protocol complete within $\tau$ slots whenever its cluster contains all $k$ participants, provided every execution is interruptible after $\tau$ slots and never fabricates records.
\end{theorem}
\begin{proof}
Run each class for $\tau$ slots, in sequence; local completion does not remove a vertex from later classes. At each class boundary, initialize every cluster protocol with its original local source records and a local clock set to zero. Execute the class's cluster protocols concurrently, with all other vertices silent. Earlier outputs are retained separately and do not change the new working states. A receiver has no neighbor transmitting for another cluster of that class, by the separation assumption. Each cluster execution is therefore identical to its isolated execution. The separate states also accommodate vertices shared by clusters in different classes.

Choose a cluster that contains the entire actual participant set. Its local protocol completes their exchange within $\tau$ slots. All participants retain its output. The $\beta$ classes take $\beta\tau$ slots in total, giving the claimed bound. The schedule need not identify in advance which cluster contains all participants.

In the exact-count case, the containing cluster has all $k$ participants and succeeds within $\tau$ slots. Interrupt every other execution after $\tau$ slots. Isolation protects the containing execution, while the monotone output buffer retains only genuine source records.
\end{proof}

The exact-count variant retains the supplied value of $k$; count gridding still requires the upper-bound correctness hypothesis of Theorem~\ref{thm:table}.

\begin{corollary}[Conditional substitutions]\label{cor:coverforms}
Under the hypotheses of Theorem~\ref{thm:cover}, with $\delta=d$ and $Q(n,k)=k\log^2n/\log k$, each indicated pair of cover parameters gives the following completion bound:
\[
\begin{array}{c|c|l}
 \alpha&\beta&\text{completion bound}\\\hline
 O(\log n)&O(\log n)&O(d\log^2n+k\log^3n/\log k)\\
 O(1)&O(\log n)&O(d\log n+k\log^3n/\log k)\\
 O(\log n)&O(1)&O(d\log n+k\log^2n/\log k)\\
 O(1)&O(1)&O(d+k\log^2n/\log k).
\end{array}
\]
\end{corollary}
\begin{proof}
Substitute the indicated values of $\alpha$, $\beta$, and $Q$ into Theorem~\ref{thm:cover}.
\end{proof}

Each row is conditional on both the indicated cover and a local protocol with the specified $Q(n,k)$; no existence claim for either is made here. Moreover, the substitution $\delta=d$ supplies the diameter parameter. Removing that knowledge requires a base family and common budgets satisfying Theorem~\ref{thm:doubling}, in addition to the hypotheses above.

\begin{proposition}\label{prop:clustercost}
For a cluster family partitioned as in Theorem~\ref{thm:cover}, write $n_U=|U|$ and $N=\max_U n_U$. If preprocessing one cluster costs $O(n_U^2\log(n_U+1))$ with a uniform implicit constant, the total cost of these computations, excluding cover construction and partitioning, satisfies
\[
 T_{\mathrm{prep}}=O\!\left(\beta nN\log(N+1)\right).
\]
\end{proposition}
\begin{proof}
A vertex belongs to at most one cluster of each class, so $\sum_U n_U\le\beta n$. Since $n_U\log(n_U+1)\le N\log(N+1)$, the sum of the individual cost bounds is at most a constant times $N\log(N+1)\sum_U n_U$.
\end{proof}

\section{Ranks and designated delivery}\label{sec:ranks}
The cluster reduction makes the local communication guarantee an explicit premise. To justify that premise through a tree-based gathering argument, one must control both progress in the tree and interference in the full radio graph. Tree ranks can organize progress, but non-tree edges may invalidate the proposed simultaneous transmissions. We first bound the ranks of the subtree formed by participant-to-root paths. We then examine the additional interference condition needed for simultaneous transmissions. Throughout this section, pruning means deleting tree branches and recomputing ranks. The counterexample concerns that operation and the specified parent-transmission schedule: the interfering radio edges remain in place.

\begin{definition}
For a rooted tree and an integer $x\ge2$, define $\rank_x(v)$ recursively. Leaves have rank one. If the largest child rank of an internal vertex is $r$, the vertex has rank $r+1$ when at least $x$ children have rank $r$, and rank $r$ otherwise.
\end{definition}

\begin{lemma}\label{lem:rank}
If the subtree rooted at $v$ has $\ell_v$ leaves, then $\ell_v\ge x^{\rank_x(v)-1}$. Consequently, the union of the root paths from $k$ selected vertices has maximum $x$-rank at most $1+\floor{\log_x k}$ when ranks are computed in that union.
\end{lemma}
\begin{proof}
Induct upward from the leaves. A leaf satisfies the assertion with equality. If a vertex inherits rank $r$, a child of rank $r$ already has at least $x^{r-1}$ leaves. If its rank increases from $r$ to $r+1$, at least $x$ disjoint child subtrees each have at least $x^{r-1}$ leaves. In either case the inequality follows.

For the second assertion, let $S$ be the selected set and let $T_S$ be the union of its root paths. Recompute ranks using only the tree edges of $T_S$. Every leaf of $T_S$ is selected: an unselected vertex on one of these paths has a successor toward the selected endpoint and hence is not a leaf. Thus $T_S$ has at most $k$ leaves. Applying the first inequality within $T_S$ gives $x^{\rank_x(v)-1}\le k$ for every vertex of that tree, proving the stated bound.
\end{proof}

Using the rank bound for communication also requires accounting for non-tree edges in the radio graph. For binary ranks, call an edge from a nonroot vertex $v$ to its parent \emph{fast} when their ranks are equal. A same-layer fast class consists of the vertices of one depth whose fast edges have one specified rank. The class has the designated-delivery property when simultaneous transmissions by its vertices, with all other vertices listening, reach all of their respective parents.

The 2-height-respecting trees of Ga\'nczorz, Jurdzi\'nski and Pelc~\cite{GJP26} impose an explicit graph-level condition: equal-height fast vertices in one layer have no common neighbor in the preceding layer. The next example identifies why pruning and recomputing ranks require rechecking this type of interference condition.

\begin{figure}[H]
\centering
\begin{tikzpicture}[every node/.style={circle,draw,minimum size=6mm,inner sep=1pt},x=1.3cm,y=.85cm]
\node (r) at (0,0) {$r$};\node (A) at (-1,-1) {$A$};\node (B) at (1,-1) {$B$};
\node (a) at (-1,-2) {$a$};\node (b) at (1,-2) {$b$};
\node (a1) at (-1.7,-3) {$a_1$};\node (a2) at (-.3,-3) {$a_2$};
\draw (r)--(A)--(a)--(a1);\draw (r)--(B)--(b);\draw (a)--(a2);\draw[dashed,thick] (B)--(a);
\end{tikzpicture}
\caption{The tree and additional radio edge used in Proposition~\ref{prop:rerank}. Pruning $a_1,a_2$ puts $a,b$ in one fast class, whose transmissions collide at $B$.}\label{fig:rank}
\end{figure}
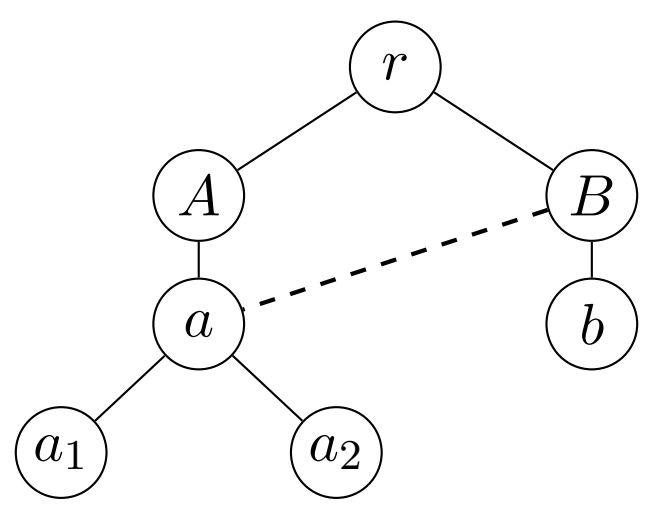

\begin{proposition}\label{prop:rerank}
Pruning a breadth-first tree to the union of selected root paths and recomputing binary ranks can destroy the designated-delivery property of same-layer fast classes, even when it held for every class before pruning.
\end{proposition}
\begin{proof}
Use the tree edges $rA,rB,Aa,Bb,aa_1,aa_2$, rooted at $r$, and include one further radio edge $Ba$. Every radio edge joins consecutive tree layers, so the tree is breadth-first. The binary ranks are
\[
 \rank_2(r)=\rank_2(A)=\rank_2(a)=2,
 \qquad \rank_2(B)=\rank_2(b)=\rank_2(a_1)=\rank_2(a_2)=1.
\]
The nonempty same-layer fast classes are the singleton classes $\{A\}$, $\{a\}$, and $\{b\}$, so each has the designated-delivery property.

Now select only $a$ and $b$ and retain their root paths. The leaves $a_1,a_2$ disappear, and the recomputed ranks of $A,a,B,b$ are all one. Thus $a$ and $b$ form a same-layer fast class. Their simultaneous transmissions collide at $B$, which is adjacent to both of them, so $b$ does not deliver to its parent. The interfering edge $Ba$ remains in the radio graph, as shown in Figure~\ref{fig:rank}.
\end{proof}

\paragraph{One-hop designated delivery}
The pruning example raises a simpler question: how many slots are sufficient when the receiving side has bounded degree? Consider a finite bipartite radio graph with sender set $A$ and receiver set $R$. Each sender $a\in A$ has a record and one designated adjacent receiver $\rho(a)\in R$. Only senders transmit, receivers always listen, and every designated receiver must obtain its sender's record directly from that sender. Senders may retry. The schedule below is constructed from the bipartite incidence graph and the designated assignments.

Define the \emph{conflict graph} $J$ on $A$ by joining distinct senders $a,a'$ whenever $a$ is adjacent to $\rho(a')$ or $a'$ is adjacent to $\rho(a)$. This graph records exactly when two designated transmissions cannot both succeed in one slot.

\begin{theorem}\label{prop:parent}
The minimum number of slots for one-hop designated delivery is $\chi(J)$, even when retries are allowed. If every receiver has at most $b\ge1$ neighboring senders, then $\chi(J)\le2b-1$, and a schedule using at most $2b-1$ slots can be constructed in polynomial time from the supplied incidence data. For every integer $b\ge1$, there is an instance with $2b-1$ senders and $2b-1$ receivers requiring exactly $2b-1$ slots.
\end{theorem}
\begin{proof}
\emph{Exact characterization.} The senders whose designated transmissions succeed in a particular slot form an independent set in $J$. If two were adjacent, one would interfere at the other's designated receiver. From any successful $T$-slot schedule, choose one successful slot for each sender. These choices give a proper $T$-coloring, so $T\ge\chi(J)$, regardless of retries. Conversely, transmitting one color class of a proper coloring in each slot makes every designated reception succeed. Thus the optimum is $\chi(J)$.

\emph{Upper bound.} Direct an interference arc from $a$ to $a'$ whenever $a\ne a'$ and $a$ is adjacent to $\rho(a')$. Each sender $a'$ has at most $b-1$ incoming arcs. Every edge of the induced conflict graph $J[U]$, for $U\subseteq A$, is represented by at least one such arc with endpoints in $U$. Hence
\[
 |E(J[U])|\le(b-1)|U|.
\]
Every nonempty induced subgraph therefore has a vertex of degree at most $2b-2$. Repeatedly remove such a vertex, then greedily color in the reverse removal order. At most $2b-1$ colors are needed. Constructing $J$ and performing this elimination takes polynomial time.

\emph{Sharpness.} Let $m=2b-1$, index senders $a_i$ and receivers $R_i$ modulo $m$, designate $a_i\to R_i$, and set
\[
 N(R_i)=\{a_i,a_{i+1},\ldots,a_{i+b-1}\}.
\]
For any distinct $i,j$, exactly one of the two cyclic differences $j-i$ and $i-j$ lies in $\{1,\ldots,b-1\}$. Thus the associated conflict graph is $K_m$. Its chromatic number is $m$, and $m$ singleton slots attain the lower bound. The case $b=1$ is immediate from a single sender and receiver.
\end{proof}

For $b=2$, the construction has three senders $a,b,c$ with designated receivers $A,B,C$ and neighborhoods $N(A)=\{a,b\}$, $N(B)=\{b,c\}$, and $N(C)=\{c,a\}$. At most one designated delivery succeeds per slot, so three slots are necessary. This recovers the three-sender obstruction and attains $2b-1$.

\section{Conclusion}
The results separate the information needed to prevent collisions from the information needed to recognize completion. A supplied distance-two coloring gives completion within $(C-1)d+1$ slots and at most $nk$ transmissions without knowing the participant set, its size, or its diameter. Geometric inputs produce explicit palettes, including $15(\Delta_0+1)$ colors from one-hop data and a constant palette for fixed separation. Without those inputs, selective schedules and independent random transmission provide alternatives under a common degree bound. The random rule has finite expected completion time, and its logarithmic cost on high-degree stars is unavoidable for that rule.

For changing subgraphs, a fixed coloring still prevents collisions, while unconditional retransmission and connectivity over bounded cycle windows ensure progress. The parameter reductions address a separate issue: how to reuse already-correct protocols when diameter guesses, table entries, or clusters are supplied. Their guarantees remain conditional on the stated base protocols and structural data. Finally, the pruning example and the exact designated-delivery theorem show why local progress measures must be accompanied by an interference argument. Improving locality-sensitive protocols beyond these bounds will require combining stronger progress mechanisms with equally explicit scheduling invariants.

\begingroup
\small
\bibliographystyle{unsrtnat}
\bibliography{references}
\endgroup
\end{document}